\documentclass[conference]{IEEEtran}

\usepackage{amsmath,amssymb,amsfonts}
\usepackage{graphicx}
\usepackage{booktabs}
\usepackage{balance}
\usepackage{cite}
\usepackage{xcolor}
\usepackage{url}

\newcommand{\jj}{\mathrm{j}}
\newcommand{\tr}{\mathrm{tr}}

\usepackage[font=footnotesize,skip=1pt]{caption}

\usepackage{enumitem}
\setlist{nosep}

\begin{document}

\title{EM-Guided Graph Learning for Fluid Antenna Beamforming under Current-Domain Constraints}

\author{
	\IEEEauthorblockN{
		Yuanhui Wu\textsuperscript{1},
		Hao Jiang\textsuperscript{2},
		Zaichen Zhang\textsuperscript{3,4}
	}
	\IEEEauthorblockA{
		\textsuperscript{1}College of Artificial Intelligence, Nanjing University of Information Science and Technology, China\\
		\textsuperscript{2}School of Cyber Science and Engineering, Southeast University, China\\
		\textsuperscript{3}National Mobile Communications Research Laboratory, Southeast University, China\\
		\textsuperscript{4}Purple Mountain Laboratories, Nanjing, China\\[-1mm]
		Emails: 202412621447@nuist.edu.cn, jiang.hao@seu.edu.cn, zczhang@seu.edu.cn
	}
}

\maketitle

\begin{abstract}
	Fluid antenna arrays (FAAs) reconfigure a finite set of radiating ports within a prescribed aperture. In compact apertures, however, channel-driven placement may cluster ports, strengthen mutual coupling, degrade radiation conditioning, increase source-voltage demand, and produce uneven current loading. This paper studies downlink multi-user beamforming with jointly optimized port placement and current-domain transmission. An electromagnetic-guided graph network predicts port layouts from channel observations and refines them using geometric and mutual-impedance information. The training objective jointly considers communication performance and electromagnetic feasibility, while a common evaluation procedure is applied to all methods. The results show that, under a common feasibility standard, the proposed method provides a controllable tradeoff among communication rate, current loading, and configuration latency.
\end{abstract}


\section{Introduction}
\label{sec:introduction}
Beamforming uses spatial degrees of freedom to enhance desired signals and suppress multi-user interference, but fixed element locations limit the response of conventional arrays. Fluid antenna systems (FASs)~\cite{wong2020bruce,wong2021fas,zhang2026fblfas} relax this constraint by reconfiguring the active radiating position within a prescribed region. Finite-blocklength performance under spatial block correlation and Cram\'er--Rao bounds for activity detection have also been characterized for FASs~\cite{zhang2026fblblock,zhang2026crbactivity}. Early studies showed that selecting a favorable port from spatially correlated channel samples can provide fading diversity~\cite{wong2020limits,chai2022port}. Later work considered broader architectures, continuous-position models, and correlation-aware analysis, including jointly correlated dual-side FASs~\cite{wong2022bruce,new2025tutorial,psomas2023continuous,zzt_pl,zhang2026dualside}. With multiple active ports, fluid antenna arrays (FAA) \cite{zhang2026finiteDesign,zhang2026finite,zhang2026emfaa} extend reconfiguration beyond favorable-port selection to control of the array geometry (a broad diversity diversity including random fading). Port coordinates affect the effective aperture, spatial sampling pattern, and array manifold, providing geometry diversity that reshapes the attainable radiation response. This freedom has been used for multi-user, near-field, and scalable beamforming~\cite{qin2023position,sumrate2024fa,chen2026nearfield,wu2026scalable}. Finite-aperture studies have analyzed the effect of port geometry on array performance~\cite{zhang2026finiteDesign,zhang2026finite}, while electromagnetic analysis has linked the geometry to mutual impedance and power, voltage, and current demands~\cite{zhang2026emfaa}. Further work has exploited array geometry for flexible beam synthesis, sidelobe suppression, and real-time port activation~\cite{xu2026flexible,liang2026psll,wu2026lbpa}.

Real-time EM-aware placement is difficult because moving a port changes the channel and circuit matrices and hence the current-domain precoder and hardware loading. Because repeating this coupled optimization for each channel is expensive, learning-based methods offer alternatives to numerical search~\cite{he2025gnnfas,xu2025practicalfas,guo2026llm}. The resulting design problem is to perform continuous-coordinate placement under a common multiport current-domain model while retaining low online complexity and circuit-consistent evaluation. Channel-aware position adaptation also relies on channel information over the aperture. Recent reconstruction methods reduce this burden by exploiting low-dimensional or geometry-dependent channel structure~\cite{wu2026lamp,liang2026reconstruction,zhang2026generative,zhang2026geometry}.

This paper develops electromagnetic-guided graph learning with current-concentration control (EMG-CC) for continuous FAA placement. EMG-CC uses the sampled channel field as node context and incorporates relative coordinates and mutual impedance into its edge messages. Training evaluates each layout with the current-domain EM-RZF rule and optimizes a loss that accounts for communication rate, geometric feasibility, hardware demand, and current concentration. The main contributions are summarized as follows:

\begin{itemize}
\item A learning formulation for continuous-coordinate placement of a finite FAA that couples a geometric multi-user channel model with a multiport current-domain evaluator based on established electromagnetic models.
\item EMG-CC, a graph coordinate generator that combines channel-field features with relative-position and mutual-impedance messages and is trained for a prescribed rate--current-loading operating point.
\item A paired-channel comparison in which all layouts undergo the same spacing projection and current-domain evaluation to characterize the tradeoff among rate, current loading, and configuration latency.
\end{itemize}

The remainder of this paper is organized as follows. Section~\ref{sec:system} presents the channel and multiport model. Section~\ref{sec:network} develops EMG-CC, Section~\ref{sec:simulation} reports the evaluation results, and Section~\ref{sec:conclusion} concludes the paper.

\emph{Notation:} Bold lowercase and uppercase letters denote vectors and matrices, respectively. $(\cdot)^T$, $(\cdot)^H$, $\tr(\cdot)$, $\|\cdot\|_F$, $\Re\{\cdot\}$, and $\Im\{\cdot\}$ denote transpose, Hermitian transpose, trace, Frobenius norm, real part, and imaginary part, respectively; $\mathcal{CN}(\boldsymbol\mu,\mathbf C)$ denotes a circularly symmetric complex Gaussian distribution.

\section{System Model}
\label{sec:system}

\subsection{Continuous Port-Placement Region}
\label{sec:placement}
Consider a downlink multi-user system where the base station has a rectangular port-placement region
\begin{equation}
	\mathcal{A}=[0,A_x]\times[0,A_y].
\end{equation}
The system activates a finite set of $M$ ports. Their coordinates are continuous-valued variables inside $\mathcal A$, collected as
\begin{equation}
	\mathbf{P}
	=
	[\mathbf{p}_1,\ldots,\mathbf{p}_M]^T,
	~
	\mathbf{p}_m=[x_m,y_m]^T\in\mathcal{A}.
\end{equation}
The finite aperture and spacing constraints are
\begin{align}
	0\le x_m\le A_x,~
	0\le y_m\le A_y,~ \forall m,\\
	\|\mathbf p_m-\mathbf p_n\|_2\ge d_{\min},~ m\ne n.
\end{align}
The spacing constraint prevents unrealistically dense port clusters, which can lead to strong mutual coupling and ill-conditioned electromagnetic matrices.

\subsection{Current-to-Received-Signal Channel}
\label{sec:channel}
For user $k$, let $\alpha_{k,\ell}$ and $\mathbf u_{k,\ell}=[u_{k,\ell},v_{k,\ell}]^T$ denote the complex gain and direction-cosine vector of path $\ell$, respectively. The element-pattern-weighted response of a current applied at port $m$ is
\begin{equation}
	b_m(\mathbf u;\mathbf P)
	=
	g_m(\mathbf u;\mathbf P)
	e^{\jj 2\pi\mathbf p_m^T\mathbf u},
\end{equation}
where $g_m$ contains the element-pattern and polarization factor. The current-to-received-signal coefficient is
\begin{equation}
	h_{k,m}(\mathbf P)
	=
	\sum_{\ell=1}^{L}\alpha_{k,\ell}
	b_m(\mathbf u_{k,\ell};\mathbf P).
\end{equation}
Stacking $h_{k,m}$ yields $\mathbf H(\mathbf P)\in\mathbb C^{K\times M}$. For mutually uncorrelated unit-power symbols, the received signal is
\begin{equation}
	y_k
	=
	\mathbf h_k^T(\mathbf P)
	\sum_{j=1}^{K}\mathbf i_j s_j+n_k,
\end{equation}
where $\mathbf h_k^T$ denotes row $k$ of $\mathbf H$, $\mathbf i_j$ is the port-current vector for stream $j$, and $n_k\sim\mathcal{CN}(0,\sigma_k^2)$. Thus, $\mathbf H\mathbf W$ is the effective multi-user channel seen by the receivers. The element factor $g_m$ can be replaced by a calibrated direction- and geometry-dependent response when such data are available.

\subsection{Current-Domain Multiport Model}
\label{sec:multiport}
For a fixed layout, the port currents are not free of circuit cost. Following standard multiport-array relations and the FAA current-domain model in~\cite{zhang2026emfaa}, the active ports interact through the position-dependent mutual-impedance matrix
\begin{equation}
	\mathbf Z(\mathbf P)\in\mathbb C^{M\times M}.
\end{equation}
Let $Z_s$ be the source impedance and $\mathbf R_{\rm loss}\succeq\mathbf0$ represent the loss resistance. In the normalization used here, the matrices
\begin{align}
	\mathbf R_{\rm acc}(\mathbf P)&=\tfrac{1}{2}(\mathbf Z+\mathbf Z^H),\\
	\mathbf R_{\rm rad}(\mathbf P)&=\mathbf R_{\rm acc}-\mathbf R_{\rm loss},\\
	\mathbf Q_v(\mathbf P)&=(\mathbf Z+Z_s\mathbf I)^H(\mathbf Z+Z_s\mathbf I).
\end{align}
associate a current vector with accepted power, radiated power, and squared source-voltage demand, respectively. In particular, for a stream-current vector $\mathbf i$, these costs are $\mathbf i^H\mathbf R_{\rm acc}\mathbf i$, $\mathbf i^H\mathbf R_{\rm rad}\mathbf i$, and $\mathbf i^H\mathbf Q_v\mathbf i$. The last expression satisfies $\mathbf i^H\mathbf Q_v\mathbf i=\|\mathbf v_s\|_2^2$, where $\mathbf v_s=(\mathbf Z+Z_s\mathbf I)\mathbf i$.

The minimum spacing is selected from the same induced-EMF model used for the mutual impedance rather than set as an arbitrary wavelength fraction. Let $r_{{\rm rad},0}=\Re\{Z_{11}\}-R_{\rm loss}$ denote the isolated-port radiation resistance. The smaller eigenvalue of the corresponding two-port radiation-resistance matrix is $r_{{\rm rad},0}-|\Re\{Z_{12}(d)\}|$. We therefore impose the pairwise screening condition
\begin{equation}
    r_{{\rm rad},0}-|\Re\{Z_{12}(d_{\min})\}|\ge \epsilon_{\rm rad}.
\end{equation}
This condition supplies a two-port local radiation-resistance margin; it does not replace the full-array eigenvalue check on $\mathbf R_{\rm rad}(\mathbf P)$.

Let
\begin{equation}
	\mathbf W=[\mathbf i_1,\ldots,\mathbf i_K]\in\mathbb C^{M\times K}
\end{equation}
denote the downlink current-domain precoder. Each column maps one unit-power data symbol to the port currents. Summing the three quadratic costs across streams gives the accepted-power, total-current, and source-voltage constraints
\begin{align}
	\tr(\mathbf W^H\mathbf R_{\rm acc}(\mathbf P)\mathbf W)
	&\le P_{\rm acc},\\
	\|\mathbf W\|_F^2
	&\le I_{\max},\\
	\tr(\mathbf W^H\mathbf Q_v(\mathbf P)\mathbf W)
	&\le V_{\max}^2 .
\end{align}
The budgets $P_{\rm acc}$, $I_{\max}$, and $V_{\max}^2$ are fixed operating limits. They are kept common to all layouts and comparison methods; their numerical values are specified only in Section~\ref{sec:simulation}. The radiation efficiency is evaluated, but is not used as an optimization target,
\begin{equation}
	\eta_{\rm rad}
	=
	\frac{\tr(\mathbf W^H\mathbf R_{\rm rad}(\mathbf P)\mathbf W)}
	{\tr(\mathbf W^H\mathbf R_{\rm acc}(\mathbf P)\mathbf W)}.
\end{equation}
We also monitor the minimum eigenvalue of $\mathbf R_{\rm rad}$ to avoid non-physical radiation matrices. For a given layout and current-domain precoder, the downlink sum rate is
\begin{equation}
	R(\mathbf P,\mathbf W)
	=
	\sum_{k=1}^{K}\log_2(1+\gamma_k).
\end{equation}
Here $\gamma_k$ is the standard multi-user SINR induced by $\mathbf H(\mathbf P)$ and $\mathbf W$. Since both the channel matrix and the circuit matrices depend on $\mathbf P$, the layout problem is strongly non-convex.

\section{EM-Guided Graph Network}
\label{sec:network}

EMG-CC is an amortized layout generator: it replaces per-channel coordinate iterations with one forward pass, while the rate and feasibility metrics are always recomputed by the current-domain evaluator in Section~\ref{sec:multiport}. It contains a channel-field encoder, an anchor-based coordinate initializer, and $S$ graph-refinement layers. The values of $S$, network width, and optimization hyperparameters are given in Section~\ref{sec:simulation} rather than treated as part of the model definition.

\subsection{Channel Observation Encoder}
\label{sec:encoder}
The input consists of channel samples on a fixed observation grid $\mathcal G=\{\bar{\mathbf p}_1,\ldots,\bar{\mathbf p}_{N_g}\}$ over the aperture. At each grid point, the real and imaginary components of the $K$-user channel samples are stacked with an average channel-magnitude map. The resulting input tensor has $2K+1$ channels and spatial dimensions $N_y\times N_x$. A lightweight convolutional encoder then extracts a latent field representation
\begin{equation}
	\mathbf F=f_{\rm enc}(\mathbf X).
\end{equation}
The encoder transforms the sampled channel field into a latent representation shared by all active ports. Bilinear sampling makes this representation available at arbitrary continuous coordinates. The observation grid is an input representation of the channel field, not a port codebook: it is used to infer the $M$ deployed coordinates, which need not coincide with grid points. Direct acquisition requires $K N_g$ complex observations per coherence interval; channel-field reconstruction is a separate problem that can reduce this burden~\cite{wu2026lamp}.

\subsection{Anchor-Based Coordinate Initialization}
\label{sec:initialization}
The $M$ active ports are initialized from deterministic anchors $\mathbf A=[\mathbf a_1,\ldots,\mathbf a_M]^T$, where $\mathbf a_m\in\mathcal A$. The anchors index the $M$ output ports and prevent the network from having to assign an arbitrary order to an unordered point set. Each anchor is encoded by its normalized coordinate, while spatial average pooling of $\mathbf F$ produces a global channel code. The initial coordinate is
\begin{equation}
	\mathbf p_m^{(0)}
	=
	\mathbf a_m
	+
	\rho\mathbf c\odot
	\tanh
	\left(
	f_{\rm init}([\mathbf e_m,\mathbf g])
	\right),
\end{equation}
where $\mathbf c$ is the anchor-cell size and $\rho$ controls the maximum initial offset. Thus, initialization is a bounded correction to a spatially distributed anchor set rather than direct regression of an unordered point cloud.

\subsection{Mutual-Impedance-Aware Refinement}
\label{sec:refinement}
After initialization, the $M$ ports form a complete directed graph and are refined by $S$ message-passing layers. At stage $s$, node $m$ represents one active port and carries its coordinate $\mathbf p_m^{(s)}$ and hidden state $\mathbf q_m^{(s)}$. For each ordered pair $(m,n)$, the edge feature is
\begin{equation}
	\mathbf e_{mn}^{(s)}
	=
	[
	\mathbf q_m^{(s)},
	\mathbf q_n^{(s)},
	\mathbf p_m^{(s)}-\mathbf p_n^{(s)},
	d_{mn}^{(s)},
	\Re\{Z_{mn}^{(s)}\},
	\Im\{Z_{mn}^{(s)}\}
	],
\end{equation}
where $d_{mn}^{(s)}=\|\mathbf p_m^{(s)}-\mathbf p_n^{(s)}\|_2$. The message received by port $m$ is
\begin{equation}
	\mathbf m_m^{(s)}
	=
	\frac{1}{M-1}
	\sum_{n\ne m} f_{\rm edge}(\mathbf e_{mn}^{(s)}).
\end{equation}
Before every graph update, bilinear interpolation samples the latent channel field at the current coordinate of each node and adds the local feature to its hidden state. A residual node update combines that state with $\mathbf m_m^{(s)}$, and a bounded two-dimensional displacement updates $\mathbf p_m^{(s)}$. The output is clipped to $\mathcal A$. This complete-graph construction is used because both spacing and mutual impedance are pairwise functions of port coordinates; a port-wise predictor would not explicitly represent how moving one port alters the circuit environment of every other port.

\subsection{Current-Domain Precoder and Loss}
\label{sec:loss}
For every predicted layout, training evaluates the rate with the same current-domain precoder used at test time. We refer to this implementation as EM-RZF. It is a circuit-weighted RZF rule, not a separately claimed precoding optimum: $\mathbf R_{\rm acc}$ penalizes accepted-power demand, $\mathbf Q_v$ penalizes source-voltage demand, and the RZF term suppresses multi-user interference.

Given a predicted layout, the current-domain EM-RZF precoder is
\begin{equation}
	\widetilde{\mathbf W}
	=
	\mathbf M^{-1}\mathbf H^H
	(\mathbf H\mathbf M^{-1}\mathbf H^H+\alpha\mathbf I)^{-1},
\end{equation}
where
\begin{equation}
	\mathbf M
	=
	\frac{\mathbf R_{\rm acc}}{\bar r}
	+\epsilon\mathbf I
	+\zeta\frac{\mathbf Q_v}{\bar q}.
\end{equation}
Here $\bar r=M^{-1}\tr(\mathbf R_{\rm acc})$ and $\bar q=M^{-1}\tr(\mathbf Q_v)$ normalize the circuit matrices so that their relative weights are not set by units alone. The non-negative constants $\epsilon$, $\zeta$, and $\alpha$ control diagonal loading, voltage weighting, and RZF regularization, respectively. The raw current matrix $\widetilde{\mathbf W}$ is scaled to meet all three current-domain budgets:
\begin{equation}
	\mathbf W=\kappa\widetilde{\mathbf W},
\end{equation}
where
\begin{align}
	\kappa
	=
	\min
	\bigg\{
	&1,\,
	\sqrt{\frac{P_{\rm acc}}
	{\tr(\widetilde{\mathbf W}^H\mathbf R_{\rm acc}\widetilde{\mathbf W})}},
	\sqrt{\frac{I_{\max}}{\|\widetilde{\mathbf W}\|_F^2}},
	\nonumber\\
	&\sqrt{\frac{V_{\max}^2}
	{\tr(\widetilde{\mathbf W}^H\mathbf Q_v\widetilde{\mathbf W})}}
	\bigg\}.
\end{align}
The scalar $\kappa$ is the most restrictive of the accepted-power, total-current, and source-voltage scales. Hence, $\mathbf W$ is feasible by construction for a fixed layout.

The loss augments the negative feasible sum rate with soft geometry and hardware-demand penalties. The spacing and radiation-conditioning penalties are
\begin{align}
S(\mathbf P)&=\operatorname{softplus}^{2}\!\left(\frac{d_{\min}-\min_{m\ne n}d_{mn}}{\tau_s}\right),\\
E(\mathbf P)&=\operatorname{softplus}^{2}\!\left(\frac{-\lambda_{\min}(\mathbf R_{\rm rad})}{\tau_e}\right),
\end{align}
where $\tau_s$ and $\tau_e$ are smoothing constants. The final-stage objective is
\begin{align}
	\mathcal L
	=
	&-R(\mathbf P,\mathbf W)
	+\eta_{\rm em}C_{\rm EM}(\mathbf P)
	+\eta_s S(\mathbf P) \nonumber\\
	&+\eta_e E(\mathbf P)
	+\eta_p B_P+\eta_i B_I+\eta_v B_V
	+\eta_{\rm cc}U(\mathbf P,\widetilde{\mathbf W}).
\end{align}
Here, $\eta_{\rm em}$, $\eta_s$, $\eta_e$, $\eta_p$, $\eta_i$, $\eta_v$, and $\eta_{\rm cc}$ are non-negative weights for the coupling, spacing, radiation-conditioning, accepted-power demand, current demand, voltage demand, and current-concentration terms, respectively.
The coupling-health term is
\begin{equation}
	C_{\rm EM}(\mathbf P)
	=
	\frac{1}{M(M-1)}
	\sum_{m\ne n}
	\frac{|Z_{mn}(\mathbf P)|^2}{|Z_{\rm ref}|^2}.
\end{equation}
The coupling term $C_{\rm EM}$ penalizes average normalized off-diagonal mutual-impedance energy. For $X\in\{P,I,V\}$, the raw-load term is $B_X=\operatorname{softplus}^{2}(\log(q_X/b_X))$, where $(q_P,q_I,q_V)$ are the accepted power, squared current norm, and squared voltage of $\widetilde{\mathbf W}$, and $(b_P,b_I,b_V)=(P_{\rm acc},I_{\max},V_{\max}^{2})$. These pre-scaling terms provide a gradient against layouts that would require substantial normalization; $\kappa$ remains the mechanism that enforces hard feasibility. The current-concentration term is
\begin{equation}
U(\mathbf P,\widetilde{\mathbf W})
	=
	\operatorname{softplus}^{2}\!\left(
\frac{\widetilde C_I(\mathbf P,\widetilde{\mathbf W})-\Gamma_{\rm cc}}{\tau_{\rm cc}}
	\right),
\end{equation}
where
\begin{equation}
C_I(\mathbf P,\mathbf W)
	=
	\frac{\max_m I_m}{\frac{1}{M}\sum_{m=1}^{M} I_m},
	~
I_m=\sum_{k=1}^{K}|i_{m,k}|^2.
\end{equation}
where $\widetilde C_I$ has the same form as $C_I$ but uses the raw EM-RZF current matrix $\widetilde{\mathbf W}$; the reported $C_I$ uses the feasible $\mathbf W$. Here, $\Gamma_{\rm cc}$ is the target and $\tau_{\rm cc}>0$ controls the transition width. Let $\mathcal L^{(s)}$ denote the loss at stage $s$. Training uses $\mathcal L_{\rm train}=\mathcal L^{(S)}+(\rho_{\rm st}/S)\sum_{s=0}^{S-1}\mathcal L^{(s)}$, where $\rho_{\rm st}$ weights the intermediate stages and $\mathcal L^{(S)}$ retains unit weight. These auxiliary terms stabilize early coordinate updates. The rate term favors channel-efficient layouts, whereas the coupling, spacing, radiation-conditioning, and raw-load penalties discourage dense, ill-conditioned, or heavily scaled layouts. The current-concentration penalty distinguishes EMG-CC from Rate-only by discouraging concentration on a few ports.

Current concentration is an implementation-oriented regularizer rather than a second link-layer utility, and it does not alter the radiated sum-rate expression. Among layouts with similar rates, it favors a more even current allocation. A lower $C_I$ is therefore not desirable if it requires a material rate loss. The parameters $\eta_{\rm cc}$ and $\Gamma_{\rm cc}$ define the selected rate--current-loading operating point, not a universal optimum or a hard bound on $C_I$.
\subsection{Inference}
\label{sec:inference}
At inference, the trained graph network predicts the coordinates in one forward pass. A deterministic collision-resolution projection removes any residual spacing violations. The current-domain evaluator then recomputes the rate, accepted power, current, voltage, radiation efficiency, minimum spacing, and current concentration for the projected layout. The network thus generates the coordinates, while the common evaluator provides all reported metrics.

\subsection{Evaluation Metrics}
\label{sec:metrics}
All methods are compared using the same post-projection current-domain evaluator. The primary communication metric is the downlink sum rate $R$. Current-loading uniformity is measured by
\begin{equation}
	C_I=\frac{\max_m I_m}{M^{-1}\sum_{m=1}^{M} I_m},
	~
	I_m=\sum_{k=1}^{K}|i_{m,k}|^2,
\end{equation}
where a smaller $C_I$ indicates a more even allocation of current across the active ports. We additionally report the accepted-power, total-current, and source-voltage feasibility rates, together with the minimum spacing and the minimum eigenvalue of $\mathbf R_{\rm rad}$. For paired test channels, $\bar R$ and $\bar C_I$ denote sample means, $R_{5\%}$ is the lower-tail 5th percentile of the sum rate, and $C_{I,95\%}$ is the upper-tail 95th percentile of current concentration. Configuration latency is measured per realization for coordinate generation; its reported median $t_{\rm med}$ excludes channel-field acquisition, coordinate actuation, and final current-domain evaluation.

\section{Simulation Results}
\label{sec:simulation}

\begin{figure*}[!t]
    \centering
    \begin{minipage}[t]{0.485\textwidth}
        \centering
        \includegraphics[width=\linewidth]{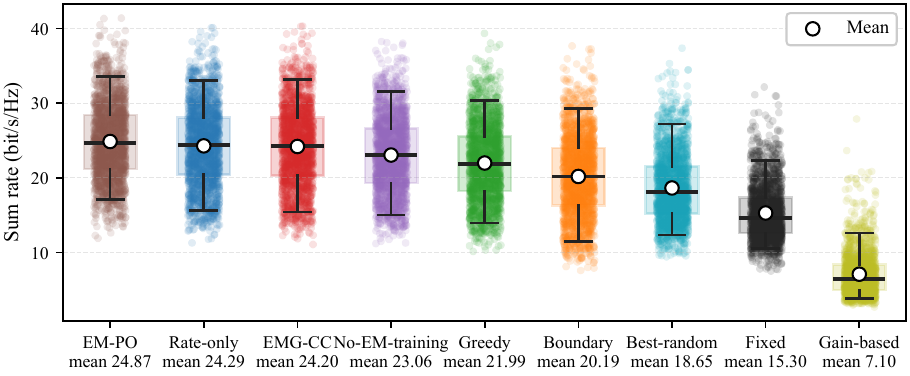}
        \captionof{figure}{Distribution of the EM-aware downlink sum rate at SNR $=10$ dB over the same $2000$ paired multi-user channel realizations. Every layout is evaluated after spacing projection by the common current-domain evaluator and its associated EM-RZF precoder. Boxes, center lines, and whiskers denote the interquartile range, median, and 5th--95th percentile interval, respectively; translucent points show individual realizations and white circles mark sample means.}
        \label{fig:rate_box}
    \end{minipage}
    \hfill
    \begin{minipage}[t]{0.485\textwidth}
        \centering
        \includegraphics[width=\linewidth]{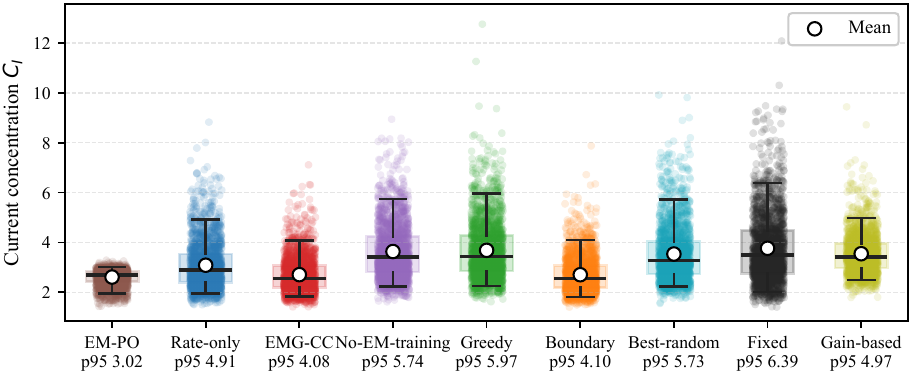}
        \captionof{figure}{Distribution of port-current concentration under the same $2000$ paired channels and final current-domain evaluation as in Fig.~\ref{fig:rate_box}. Here $C_I=\max_m I_m/\bar I$, where $I_m=\sum_k|i_{m,k}|^2$ is the current power at port $m$ and $\bar I$ is its average across the $M$ active ports. Lower $C_I$ indicates a more even current allocation. Boxes, center lines, whiskers, points, and white circles use the same statistical conventions as Fig.~\ref{fig:rate_box}.}
        \label{fig:cc_box}
    \end{minipage}
\end{figure*}

\paragraph{Setup}
\label{sec:setup}
The main experiment uses a $4\lambda\times2\lambda$ aperture with $M=32$ active ports, $K=8$ users, and $L=4$ paths per user. Channel samples are provided on a $16\times16$ observation grid with $g_m=1$, nominal user directions uniformly spanning $[-55^\circ,55^\circ]$, Gaussian path-angle offsets of $12^\circ$, an exponentially decaying path-power profile, and user ranges uniformly sampled from $60\lambda$ to $200\lambda$, which exceed the $1.5$-Fraunhofer-distance margin. Unless otherwise stated, the SNR is $10$ dB. The induced-EMF model uses identical parallel center-fed thin half-wave dipoles in free space~\cite{zhang2026emfaa}, with $Z_{11}=74.1+\jj42.5\,\Omega$, $R_{\rm loss}=1\,\Omega$, and $r_{{\rm rad},0}=73.1\,\Omega$. The same model determines $Z_{12}(d)$ and, with $\epsilon_{\rm rad}=9\,\Omega$, gives $d_{\min}=0.125273\lambda$ through the screening rule in Section~\ref{sec:multiport}. This is a normalized thin-dipole multiport model rather than a device-specific impedance calibration. The operating limits are $P_{\rm acc}=1$, $I_{\max}=0.03621$, and $V_{\max}^2=443.4$; they impose common accepted-power, current, and voltage constraints and are not transmitter-chain ratings. The limits are fixed before evaluation and enforced by the same scaling rule for all methods.

The network uses 64-dimensional node states and $S=5$ refinement stages. The training, validation, and test sets contain 8000, 1600, and 2000 disjoint realizations, generated with separate deterministic seeds. AdamW is run for 100 epochs with learning rate $1.5\times10^{-3}$, weight decay $10^{-4}$, and batch size 16; the checkpoint with the lowest validation loss is retained. EM-RZF uses $(\alpha,\epsilon,\zeta)=(0.1,0.05,0.02)$. The loss weights are $(\eta_{\rm em},\eta_s,\eta_e,\eta_{\rm cc})=(0.1,12,0.1,0.1)$ and $\eta_p=\eta_i=\eta_v=0.03$, with $(\tau_s,\tau_e,\tau_{\rm cc},\rho_{\rm st})=(0.02,2,0.25,0.15)$. Validation among $(\Gamma_{\rm cc},\eta_{\rm cc})\in\{(4,0.1),(3,0.1),(3,0.3)\}$ selects $(\Gamma_{\rm cc},\eta_{\rm cc})=(3,0.1)$. The activation frequencies reported in Section~\ref{sec:setup} confirm that none of the three operating limits is redundant.

For an apples-to-apples comparison, all methods use $M=32$ active ports and are evaluated after the same projection and current-domain evaluator. The compared methods are:
\begin{itemize}\setlength{\itemsep}{0pt}\setlength{\parsep}{0pt}\setlength{\topsep}{2pt}
	\item \textbf{\textit{EMG-CC}}: the proposed mutual-impedance-aware graph network trained with the complete current-domain objective.
	\item \textbf{\textit{EM-PO}}: a per-realization projected coordinate optimizer initialized by the uniform anchors and updated for 200 Adam steps using the same EM-RZF objective, followed by collision projection.
	\item \textbf{\textit{Rate-only}}: the same graph, impedance-edge features, and EM-RZF evaluator as EMG-CC, trained with only the EM-aware rate and common spacing terms.
	\item \textbf{\textit{No-EM-training}}: the same coordinate backbone trained with conventional RZF and Euclidean current normalization; its graph edges and loss omit mutual-impedance and accepted-power, radiation, voltage, coupling, and current-concentration information, while spacing remains active.
	\item \textbf{\textit{Fixed}}: a deterministic uniform in-aperture grid.
	\item \textbf{\textit{Boundary}}: a channel-independent perimeter layout.
	\item \textbf{\textit{Best-random}}: the highest-rate layout among eight independently sampled spacing-feasible candidates.
	\item \textbf{\textit{Gain-based}}: the observation-grid points with the largest aggregate multi-user channel power.
	\item \textbf{\textit{Greedy}}: a channel-adaptive selector that sequentially adds the spacing-feasible point maximizing a regularized multi-user log-determinant.
\end{itemize}
Rate-only isolates the effect of the additional EM penalties while retaining current-domain rate evaluation and hard scaling. No-EM-training tests the value of circuit information during graph refinement and training-time precoding. At test time, all nine methods use the same post-projection current-domain evaluator, preventing rate gains from being attributed solely to a less restrictive training-time power convention.

\paragraph{Main Rate Comparison}
\label{sec:rate}
Fig.~\ref{fig:rate_box} compares sum-rate distributions over paired test channels. EM-PO is the optimization-quality reference, while EMG-CC amortizes placement into one graph-network pass. Table~\ref{tab:comparison} shows that EM-PO achieves the strongest rate and current-loading statistics but incurs much larger online configuration latency. EMG-CC therefore serves as a low-latency approximation to iterative EM placement rather than a replacement for its final optimization quality.

Fig.~\ref{fig:cc_box} gives the corresponding current-loading distributions. Relative to Rate-only, EMG-CC reduces both the mean and upper-tail concentration, with the corresponding rate change reported in Table~\ref{tab:comparison}. Together with EM-PO, the results expose a three-way tradeoff among rate, current loading, and configuration latency.

\begin{table}[t]
\centering
\caption{Common-evaluator comparison at 10 dB on 2000 paired channels. $R_{5\%}$ and $C_{I,95\%}$ are channel percentiles; $t_{\rm med}$ excludes CSI acquisition and final evaluation.}
\label{tab:comparison}
\footnotesize
\setlength{\tabcolsep}{0pt}
\begin{tabular*}{\columnwidth}{@{\extracolsep{\fill}}lrrrrr@{}}
\toprule
Method & $\bar R$ & $R_{5\%}$ & $\bar C_I$ & $C_{I,95\%}$ & $t_{\rm med}$ (ms)\\
\midrule
EM-PO            & \textbf{24.871} & \textbf{17.069} & \textbf{2.613} & \textbf{3.018} & 89.86\\
\textbf{EMG-CC} & 24.203 & 15.434 & 2.708 & 4.083 & 2.79\\
Rate-only        & 24.285 & 15.624 & 3.088 & 4.914 & 2.55\\
No-EM-training   & 23.058 & 15.026 & 3.637 & 5.743 & 2.53\\
Boundary         & 20.190 & 11.451 & 2.706 & 4.099 & 0.02\\
Best-random      & 18.648 & 12.354 & 3.537 & 5.729 & 93.36\\
Greedy           & 21.990 & 13.963 & 3.685 & 5.974 & 4.14\\
Fixed            & 15.296 & 10.554 & 3.767 & 6.394 & 0.02\\
Gain-based       & 7.102  & 3.861  & 3.549 & 4.975 & 0.80\\
\bottomrule
\end{tabular*}
\end{table}

\begin{figure}[t]
    \centering
    \includegraphics[width=\linewidth]{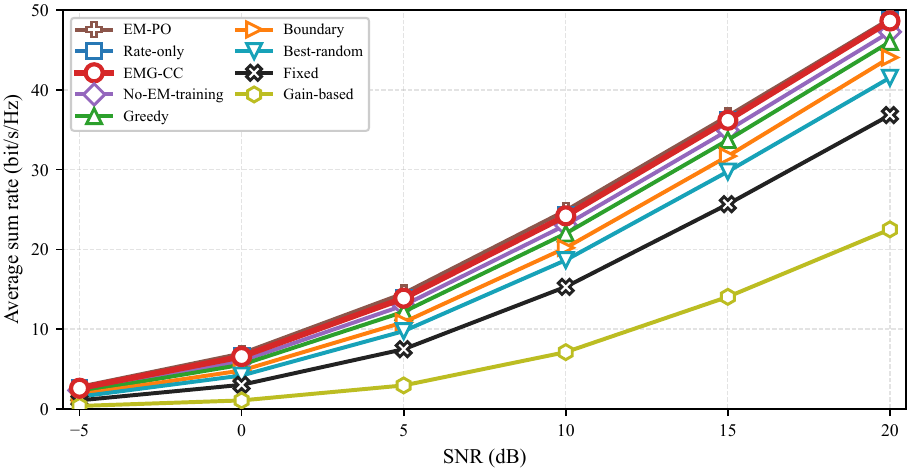}
    \caption{Mean EM-aware downlink sum rate versus SNR over $2000$ paired multi-user channel realizations. For each method and realization, the saved port layout and its current-domain precoder are held fixed while the receiver-noise level is swept from $-5$ to $20$ dB. EM-PO is the 200-step iterative reference; the remaining layouts are generated once per realization. The common evaluator therefore isolates SNR dependence from a change in port placement.}
    \label{fig:sweep_mean}
\end{figure}

\paragraph{SNR Sweep}
\label{sec:snr}
Fig.~\ref{fig:sweep_mean} reports the mean sum rate over the evaluated SNR range. EM-PO provides the rate-quality reference across the sweep, while EMG-CC remains close to the learned Rate-only variant and exceeds the remaining feedforward and heuristic methods. The figure does not establish behavior outside the simulated channel and SNR ranges.

\begin{figure}[t]
    \centering
    \includegraphics[width=\columnwidth]{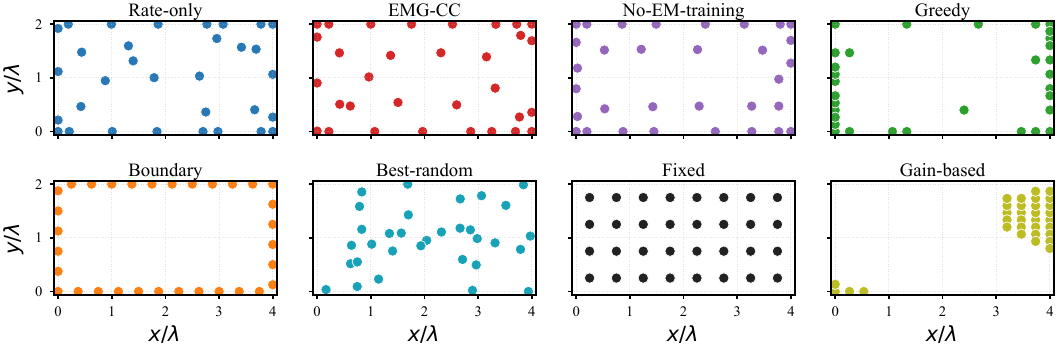}
    \caption{Representative active-port layouts for one common test-channel realization within the $4\lambda\times2\lambda$ continuous placement region. Each panel contains the finite deployed set of $M=32$ active ports. Fixed and Boundary are channel independent; the remaining methods adapt the layout to the same observed channel field. The realization was selected by its joint median distance over rate and current concentration across methods, rather than by the performance of any individual method.}
    \label{fig:ports}
\end{figure}

\paragraph{Representative Port Layouts}
\label{sec:layouts}
Fig.~\ref{fig:ports} shows one representative channel realization. Fixed and Boundary are channel independent. Gain-based clusters ports in locally strong grid regions, whereas Greedy and the learned methods account for multi-user utility. EM-PO is omitted from this eight-panel visualization because it is the iterative quality reference and the figure is intended to compare the one-shot and heuristic layout generators. These panels explain geometry but are not statistical evidence; the paired distributions in Figs.~\ref{fig:rate_box}--\ref{fig:cc_box} provide that comparison.

\paragraph{Discussion}
\label{sec:discussion}
EMG-CC predicts the port coordinates in one pass, and all methods are evaluated after the same spacing projection using current-domain EM-RZF. The model includes layout-dependent circuit and radiation costs but omits full-wave element-pattern changes, substrates, packaging, actuator errors, and port quantization. EMG-CC satisfies the spacing constraint even before projection. Accepted power, current, and voltage set the final scale in $50.7\%$, $13.9\%$, and $35.4\%$ of the test cases, respectively; these are normalized constraints rather than device-level ratings.

\section{Conclusion}
\label{sec:conclusion}
EMG-CC learns continuous FAA coordinates with a multiport current-domain evaluator, and every layout is tested under the same feasibility rule. On the evaluated channels, one graph-network pass approaches EM-PO in rate and current loading without its per-channel iterative search; Rate-only and No-EM-training separate the effects of current-concentration control and circuit information. The results are limited to the normalized far-field isotropic-element model; device-level validation requires calibrated element responses, port quantization, actuator dynamics, and pilot-efficient channel-field acquisition.

\balance

\end{document}